\documentclass[12pt]{article}
\pdfoutput=1

\usepackage[nosort]{cite}
\usepackage{epsfig}
\usepackage{amsmath}
\usepackage{amsfonts}
\usepackage{amssymb}
\usepackage{multirow}
\usepackage{array}
\usepackage{subfigure}

\usepackage[affil-it]{authblk}

\title{\bf Vector resonances in weak-boson-fusion at future pp colliders }

\author[1]{Kirtimaan Mohan}
\author[1, 2]{ Natascia Vignaroli}
\affil[1]{\small Department of Physics and Astronomy, Michigan State University, East Lansing 48824, USA}
\affil[2]{\small CP3-Origins \& DIAS, University of Southern Denmark, Campusvej 55, 5230 Odense M, Denmark}

\date{} 

\begin{document}

\maketitle

\begin{abstract}
We present a first estimate of the reach of future pp colliders, the 14 TeV LHC and a futuristic 100 TeV pp collider, on a vector resonance, specifically a $W^{'}$, produced via weak-boson-fusion, and decaying dominantly into $tb$. The analysis is motivated by Composite Higgs, Randall-Sundrum and Little Higgs scenarios, which predict the existence of vector resonances with a large coupling to $W$ and $Z$ longitudinal bosons.
In particular, in composite Higgs models with partial compositeness, the standard Drell-Yan production channel is suppressed at large coupling while the weak-boson-fusion is enhanced and could thus provide a unique opportunity to directly test the large coupling regime of the theory. We outline a search strategy for the $W^{'}$ in the weak-boson-fusion channel and present the reach of future colliders on the $W^{'}$ mass {\it vs} coupling parameter space.
\end{abstract}

\newpage

\section{Introduction}
Vector resonances $V$ are a prediction of many beyond the Standard Model (BSM) theories. In compelling scenarios to address the hierarchy problem, as minimal composite Higgs models (MCHM) \cite{Agashe:2004rs} and Little Higgs \cite{ArkaniHamed:2002qx}, the Higgs is a pseudo Nambu-Goldstone boson (pGB) associated with a global symmetry of a new strongly-interacting sector which triggers electroweak-symmetry-breaking (EWSB). Vector resonances, which emerge from the new strong dynamics, are predicted to interact strongly with the composite Higgs and would-be Goldstone bosons, and thus to longitudinal $W_L, Z_L$ bosons. On the other hand, in general, vector resonances are not expected to couple strongly to light quarks.\\
In the scenario of partial compositeness \cite{Kaplan:1991dc}, the hierarchy among the Standard Model (SM) fermion masses is naturally explained through a seesaw-like mechanism. In particular, the top and bottom masses can be naturally generated through their sizable mixing with the strong sector without being in conflict with flavour observables \cite{Contino:2010rs, KerenZur:2012fr, Cacciapaglia:2015dsa}. Third generation quarks are thus strongly coupled to $V$ but light quarks are predicted to be weakly coupled to the vector resonances, with an interaction that is inversely proportional to the $V$ coupling to $W_L/Z_L$, $g_V$ \cite{Contino:2006nn, Vignaroli:2014bpa, Pappadopulo:2014qza}. This implies that in more strongly-coupled scenarios of the strong electroweak sector, $1\ll g_V < 4 \pi$, the production of vector resonances via Drell-Yan (DY) is suppressed (by $\sim 1/g^2_V$) and less sensitive to the discovery of $V$. On the other hand the alternative vector-boson-fusion (VBF) production is enhanced (by $\sim g^2_V$) and can thus be used to directly probe a strongly-coupled (but still perturbative) regime that could be otherwise difficult to test via the DY channel \cite{Thamm:2015zwa}.\\
In this paper we will study the reach of future pp circular colliders on vector resonances produced by VBF. In particular, motivated by partial compositeness, which as anticipated predict a large $V$ coupling to third generation quarks, we will focus on the channel $W^{'} \to tb$. \\
VBF can be particularly powerful, due to its $t$-channel nature, at a futuristic 100 TeV pp collider. We will therefore analyze the $W^{'} \to tb$ channel in VBF at both the high luminosity LHC with $\sqrt{s}=14$ TeV and at a futuristic 100 TeV collider. \\
Recent studies for the reach of a 100 TeV collider on vectorlike quarks, which are also a general prediction of composite Higgs, Randall-Sundrum and Little Higgs models, have been presented in \cite{Agashe:2013hma, Andeen:2013zca}, while the reach for different dijet vector resonances in the DY channel has been estimated in \cite{Yu:2013wta}. Other recent studies focused on the possibility to better explore the EWSB sector in VB scattering at the LHC \cite{Englert:2015oga} 
and at a 100 TeV collider \cite{Degrande:2013yda} or in double-Higgs production \cite{Azatov:2015oxa, Kotwal:2015rba}. \\

The rest of the paper is organized as follows: we will give a brief overview of the model and describe the  phenomenology of vector resonances in sec. \ref{sec:model}. We present our 
search strategy in sec. \ref{sec:search} and our results, showing the reach of future pp colliders on the $W^{'}$ mass {\it versus} coupling parameter space in sec. \ref{sec:results}. We draw our conclusion in sec. \ref{sec:conclusions}.

\section{The model}\label{sec:model}

We will consider the two-site description derived in \cite{Contino:2006nn}, which captures the relevant phenomenology of MCHM \cite{Agashe:2004rs}\footnote{The chiral Lagrangian up to $p^4$ order for MCHM has been derived in \cite{Contino:2011np} (recently Ref. \cite{Yepes:2015zoa} derived it for a general non-linear left-right dynamical Higgs theory). }. 
The model comprises a strongly interacting sector, made up of particles which become composite at the TeV scale and a weakly coupled sector of elementary states with gauge symmetries analogous to the SM.
SM particles and new heavy resonances emerge from the mixing of these two sectors.
We will refer to MCHM with the minimal coset $SO(5)/SO(4)$ \footnote{This is a minimal coset which includes a custodial symmetry protection to the $\rho$ parameter.}, where the composite sector possesses a gauged $SU(2)^{comp}_L \times SU(2)^{comp}_R\times U(1)_X$ symmetry. The hypercharge is realized as $Y=T^3_R+X$.
The gauge bosons associated with the $SU(2)^{comp}_L$ symmetry in the composite sector \footnote{In our analysis we are interested in the phenomenology of the $SU(2)^{comp}_L$ vector resonances and we neglect the vector resonances of $SU(2)^{comp}_R\times U(1)_X$. The phenomenology of the $W^{'}_R$ associated to $SU(2)^{comp}_R$ has been briefly described in \cite{Vignaroli:2014bpa}. }, $W^{*}_{\mu}$, mix, through a mass-mixing term, with the $W_{\mu}$ of the $SU(2)^{ele}_L$ in the elementary sector. After diagonalizing the mixing, the eigenstates include the SM $W$ boson, which will become massive after the EWSB (and the SM $W_3$, which will become part of the SM $Z$ boson) and the new heavy vector resonances $W^{' \pm}=\frac{W^{*}_{1}\mp i W^{*}_{2}}{\sqrt{2}}$ and $Z^{'}=W^{*}_3$. The rotation from the elementary-composite basis to the eigenstate basis is determined by

\begin{equation}\label{eq:ct2}
\cot\theta_2=\frac{g^{*}_2}{g^{el}_{2}} \qquad g_2=g^{el}_{2}\cos\theta_2=g^{*}_{2}\sin\theta_2 \ ,
\end{equation}   
\noindent
where $g_2=e/\sin\theta_W$ is the SM gauge coupling and $g^{*}_2$ and $g^{el}_{2}$ are respectively the $SU(2)^{comp}_L$ and the $SU(2)^{ele}_L$ couplings \footnote{$SU(2)^{comp}_L$ is a broken gauge symmetry, the associated vector resonances possess a bare mass $M^{*}_V$. After the mass-mixing diagonalization, $SU(2)^{comp}_L \times SU(2)^{ele}_L $ breaks down to the SM $SU(2)_L$ and the physical mass of the vector resonances is given by $M_{V}=M^{*}_V/\cos\theta_2$. After the EWSB, $W^{'}$ and $Z^{'}$ masses receive corrections from this value coming from electroweak mixing effects.}. \\
Akin to mixing in the bosonic sector, elementary quarks in the weakly-coupled sector mix with composite fermionic partners in the strongly-coupled sector through linear mass-mixing terms \cite{Kaplan:1991dc}. After diagonalizing the mixing, we have a scenario of partial compositeness of the SM quarks, which become admixtures of their elementary and composite modes and acquire their masses through the interactions of their composite modes with the composite Higgs. Heavier quarks, such as the top and the bottom, have thus a sizable degree of compositeness, while, as anticipated, light quarks have a negligible composite component. In particular, the top mass is generated as:
\begin{equation}\label{eq:mt}
M_{top} \simeq Y_{*} \frac{v}{\sqrt{2}} s_L s_R \ ,
\end{equation}
where $v=246$ GeV is the electroweak scale, $Y_{*}$ is the Yukawa coupling between the composite Higgs and the composite top-partners, and $s_{L} (s_R)$ represents the degree of compositeness of the left-handed (right-handed) top. $SU(2)_L$ gauge invariance implies the same $s_L$ degree of compositeness for $t_L$ and $b_L$. $c_L=\sqrt{1-s^2_L}$ parametrizes the superposition of the $(t_L, b_L)$ doublet with the elementary state $(t^{ele}_L, b^{ele}_L)$.\\ 

The Lagrangian describing the interactions of vector resonances with SM bosons and fermions reads:   

\begin{align}\label{eq:LV}
\begin{split}
\mathcal{L}_V  = & -g_2 M_{W} \cot\theta_2 W^{'+}_{\mu}W^{-\mu}h  -\frac{g_2}{c_W} M_{W} \cot\theta_2 Z^{'}_{\mu}Z^{\mu}h \\
&+ i\frac{g_2}{c_W}\cot\theta_2 \frac{M^2_W}{M^2_{W'}} \left[ Z^{\mu} W^{+\nu } \left(\partial_\mu W^{' -}_{\nu} -\partial_\nu W^{'-}_{\mu} \right) \right.\\
& + Z^{\mu} W^{'  +\nu} \left(\partial_\mu W^{ -}_{\nu} -\partial_\nu W^{ -}_{\mu} \right) + W^{' + \mu} W^{ - \nu } \left(\partial_\mu Z_{\nu} -\partial_\nu Z_{\mu} \right) \left. \right. \big ] \\
&+ ig_2\cot\theta_2 \frac{M^2_W}{M^2_{Z'}} \left[ W^{+\mu} W^{-\nu } \left(\partial_\mu Z^{' }_{\nu} -\partial_\nu Z^{'}_{\mu} \right) \right.\\
& - W^{+ \mu} Z^{' \nu} \left(\partial_\mu W^{ -}_{\nu} -\partial_\nu W^{ -}_{\mu} \right) - Z^{' \mu} W^{ - \nu } \left(\partial_\mu W^{+}_{\nu} -\partial_\nu W^{+}_{\mu} \right) \left. \right. \big ] \\
&- \frac{g_2}{\sqrt{2}} \tan\theta_2 W^{' +}_{\mu}\left( \bar{q}^u_L \gamma^{\mu} q^{d}_L + \bar{\nu}_{lL} \gamma^{\mu} l^{-}_L \right)\\
&+ \frac{g_2}{\sqrt{2}}  W^{' +}_{\mu}\left( \bar{t}_L \gamma^{\mu} b_L\right)\left( s^2_L \cot\theta_2-c^2_L \tan\theta_2\right)\\
&- g_2 \tan\theta_2 Z^{' }_{\mu}\left( \bar{q}_L \gamma^{\mu} \tau^3 q_L + \bar{\nu}_{lL} \gamma^{\mu} \nu_{lL} - l^{+}_L\gamma^{\mu} l^{-}_L \right)\\
&+ g_2 Z^{' }_{\mu}\left( \bar{t}_L \gamma^{\mu} t_L - \bar{b}_L \gamma^{\mu} b_L\right)\left( s^2_L \cot\theta_2-c^2_L \tan\theta_2\right) + \text{H. c.}
\end{split}
\end{align}
where $s_W(c_W) \equiv \sin\theta_W (\cos\theta_W)$ and $q=(q^u, q^d)$ represents a doublet of the first or the second generation of quarks. \\

\begin{figure}
\centering
\includegraphics[width=0.6\textwidth]{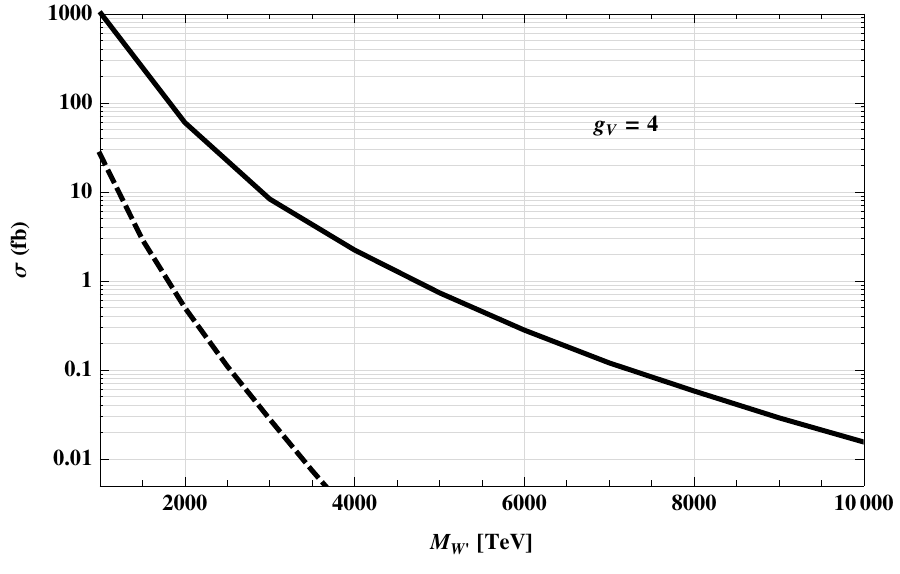} \\[0.2cm]
\includegraphics[width=0.45\textwidth]{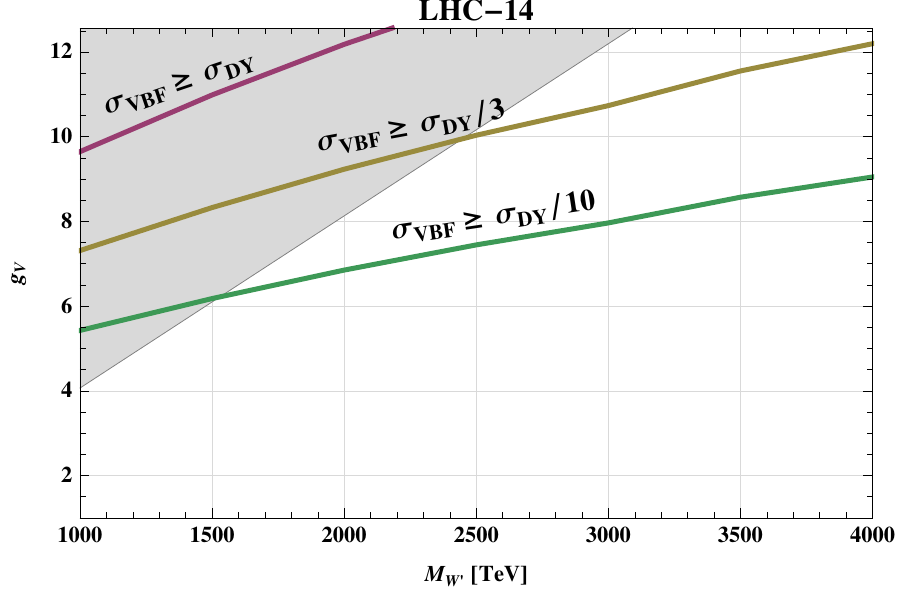} 
\includegraphics[width=0.45\textwidth]{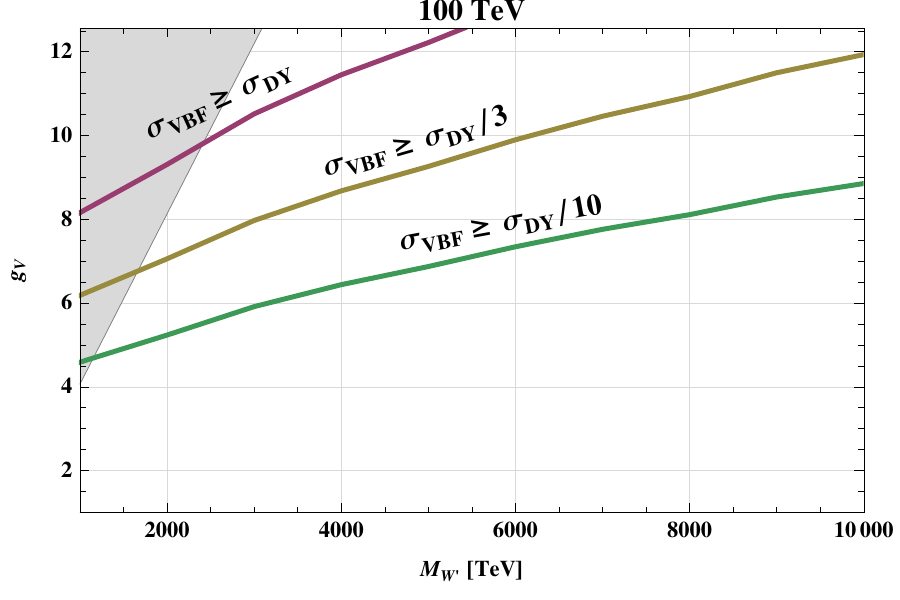} 
\caption{\small Upper Plot: cross section for the $W^{'}$ VBF production at the LHC-14 (dashed curve) and at a futuristic 100 TeV pp collider (thick curve) for a coupling $g_V=4$. Cross sections scale as $g^2_V$ with the coupling. We have applied a 30 GeV cut on the jet $p_T$ and a rapidity acceptance $|\eta_j|<5$. Lower plots: contours of different ratios of the VBF over DY $W^{'}$ production cross sections on the ($M_{W^{'}}$, $g_V$) parameter space at the LHC-14 (left plot) and at a 100 TeV collider (right plot). The shaded areas in the upper-left corner of the parameter space correspond to values $g_V v/M_{V} > 1$ which are indicative of a theoretically excluded region (where $v/f\gtrsim 1$) in MCHM.}
\label{fig:xsec}
\end{figure}

We see that the $\theta_2$ angle in (\ref{eq:ct2}) controls the interactions of the vector resonances. In particular, the $W^{'}/Z^{'}$ coupling to light quarks, which, in the partial compositeness scenario, have negligible mixings with the composite sector and are thus completely elementary states, is given by $g_2 \tan\theta_2$, while the coupling to composite modes, as the longitudinal $W/Z$ bosons, is given by:  
\begin{equation}\label{eq:gV}
g_V = g_2 \cot\theta_2 \ .
\end{equation}   
This implies that the VBF production of the vector resonances is controlled by $g^2_V$ and is thus enhanced in the regime of more strongly-coupled electroweak sectors, while the DY is controlled by $g^2_2/g^2_V$ and is thus suppressed for large $g_V$ couplings.\\
The two lower plots in Fig. \ref{fig:xsec} show contour regions in the mass-coupling parameter space with different values of the ratio between the VBF and the DY production cross sections for a $W^{'}$ resonance. We see that VBF production gets a cross section of the same order of magnitude of the DY for $g_V \gtrsim 6$. In this large-coupling regime, VBF becomes therefore a dominant production mechanism. Further, one can take advantage of the unique topology of the VBF mechanism. The VBF signal is characterized by the presence of two forward-backward final jets that permit a clear distinction of the signal from the background even in the case of broad vector resonances, a regime which is instead difficult to explore via the DY production \cite{Pappadopulo:2014qza}. VBF is thus a promising production mechanism to analyze at future colliders that can give complementary information to those from the searches in the DY channel and possibly give access to the large $g_V$ coupling regime.\\
The shaded region in the upper-left corner of the vector resonance parameter space corresponds to values $g_V v/M_{V} > 1$. In MHCM with a pGB Higgs, this region is indicative, up to $\mathcal{O}(1)$ corrections
, of a theoretically forbidden parameter space where $v/f \simeq g_V v/M_{V} >1$, with $f$ representing the compositeness scale \cite{Giudice:2007fh}.  
The upper panel in Fig. \ref{fig:xsec} shows the cross section for the VBF production of a $W^{'}$ resonance ($W^{'+}+W^{'-}$) at the 14 TeV LHC (LHC-14) and at a futuristic 100 TeV collider, for a fixed coupling $g_V=4$. It is evident from the plot that the VBF production takes considerable advantage of the increase in the collider center-of-mass-energy. At a futuristic 100 TeV collider the VBF yield is significant and could allow to access the multi-TeV mass region. The plots in Fig. \ref{fig:xsec} have been obtained by applying an acceptance cuts $|\eta|<5$ on the rapidity of the VBF jets. As we will show in the next section, the VBF sensitivity would be greatly increased if this acceptance could be enlarged at a futuristic collider.\\

\begin{figure}
\centering
\includegraphics[width=0.45\textwidth]{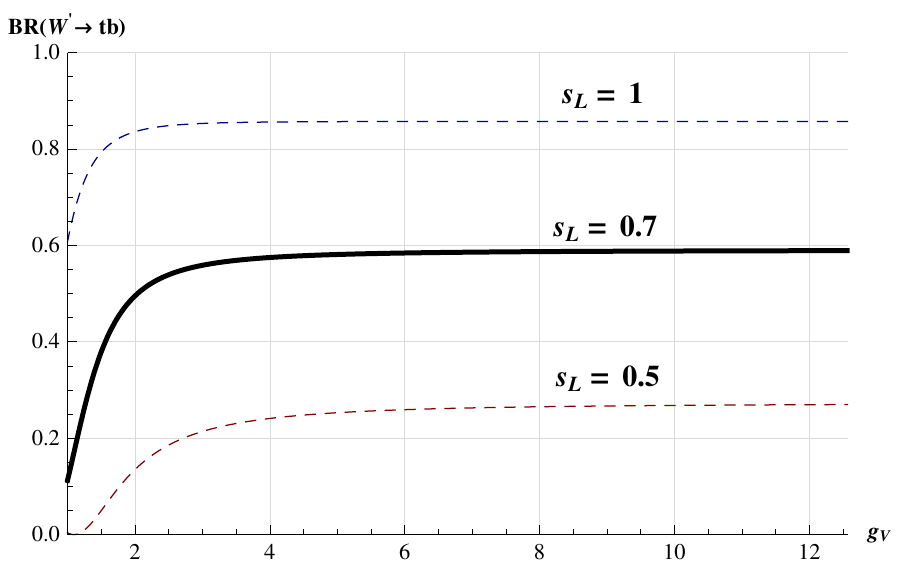} 
\includegraphics[width=0.45\textwidth]{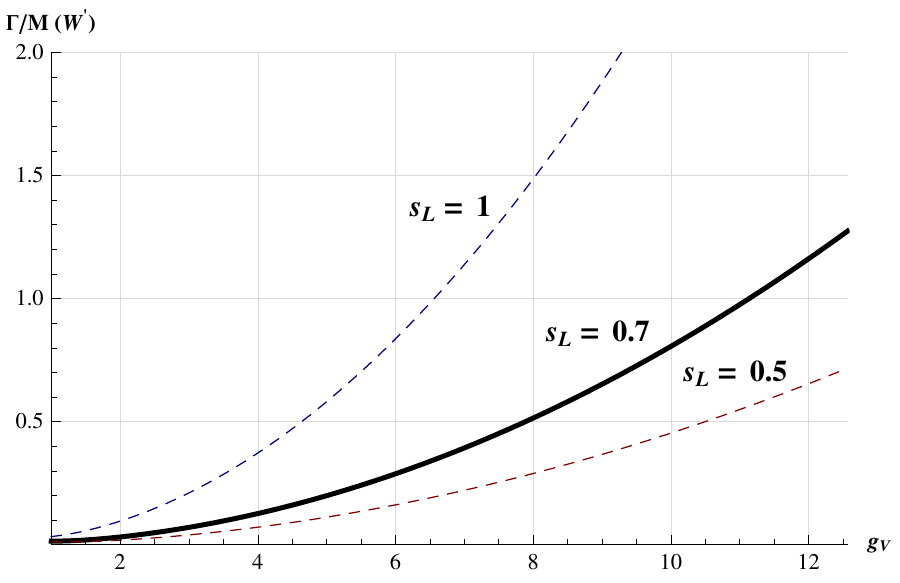} 
\caption{\small $W^{'} \to tb$ branching ratio (left plot) and $W^{'}$ width-over-mass ratio (right plot) as functions of the $g_V$ coupling for different top degrees of compositeness ($s_L$ values). }
\label{fig:BR-width}
\end{figure}

In our analysis we will focus on the VBF production of a $W^{'}$ resonance.
The $W^{'}$ decay rates are the following (we refer the reader to Ref. \cite{Vignaroli:2014bpa} for more details on the $W^{'}$ phenomenology): 

\begin{align}\label{eq:decays}
\begin{split}
 \Gamma(W^{'+}\to W^{+}_L Z_L)& = \Gamma(W^{'+}\to W^{+}_L h)  =\frac{g^2_2}{192 \pi}M_{W^{'}} \cot^2\theta_2\\
\Gamma(W^{'+}\to l^{+}\nu) & =\frac{g^2_2}{48 \pi}M_{W^{'}} \tan^2\theta_2 \\
 \Gamma(W^{'+}\to \bar{q}q') & =\frac{g^2_2}{16 \pi}M_{W^{'}} \tan^2\theta_2 \\
  \Gamma(W^{'+}\to t\bar{b}) &=\frac{g^2_2}{16 \pi}M_{W^{'}} \left(s^2_L \cot\theta_2 - c^2_L \tan\theta_2 \right)^2 
\end{split}
\end{align}
Motivated by partial compositeness, we will consider a moderately large, $s_L=0.7$, degree of compositeness for the third generation quarks. We will thus focus our analysis on the $W^{'}\to tb$ channel. Similar search strategy and sensitivity are expected for $Z^{'} \to t\bar{t}$.\\
For small values of the top degree of compositeness ($s_L\ll 0.5$ values imply $BR(W^{'}\to tb)\ll 0.2$, as shown in Fig. \ref{fig:BR-width}), which are however less natural in the partial compositeness scenario, more promising channels to analyze could be $W^{'} \to WZ/Wh$. We leave the study of these channels to a future work \cite{WZ}  \footnote{ For the theoretical scenarios we are considering in our study, the leptonic decays of the vector resonances are strongly suppressed (see, for example, \cite{Vignaroli:2014bpa}) and will be thus overlooked in our analysis. The leptonic channels have been analyzed in \cite{deBlas:2012qp}, on searches for $W^{'}$ and $Z^{'}$ resonances in a model-independent framework.}. Here we are mainly interested in estimating the sensitivity of the VBF production mechanism, regardless of the specific decay channel. We thus assume, for simplicity, that vectorlike quarks are heavier than vector resonances. In a more natural scenario with vectorlike quarks at the $\sim$1 TeV scale, other promising channels to analyze in VBF are those of vector resonance decays to top partners \cite{Vignaroli:2014bpa}. \\
For $s_L=0.7$, the $W^{'} \to tb$ branching ratio is of about 0.6 in the more strongly-coupled regime $g_V \gtrsim 3$, relevant to our analysis (Fig. \ref{fig:BR-width}). Fig. \ref{fig:BR-width} shows the $tb$ branching ratio (left plot) and the width-over-mass ratio (right plot) for the $W^{'}$. For $s_L=0.7$, the $W^{'}$ becomes a broad resonance, $\Gamma/M\gtrsim 0.3$, for $g_V \gtrsim 6$. In our study we will analyze both the narrow-width and the broad-width regimes.

\section{Monte Carlo simulation and search strategy}\label{sec:search}

\begin{figure}
\centering
\includegraphics[width=0.6\textwidth]{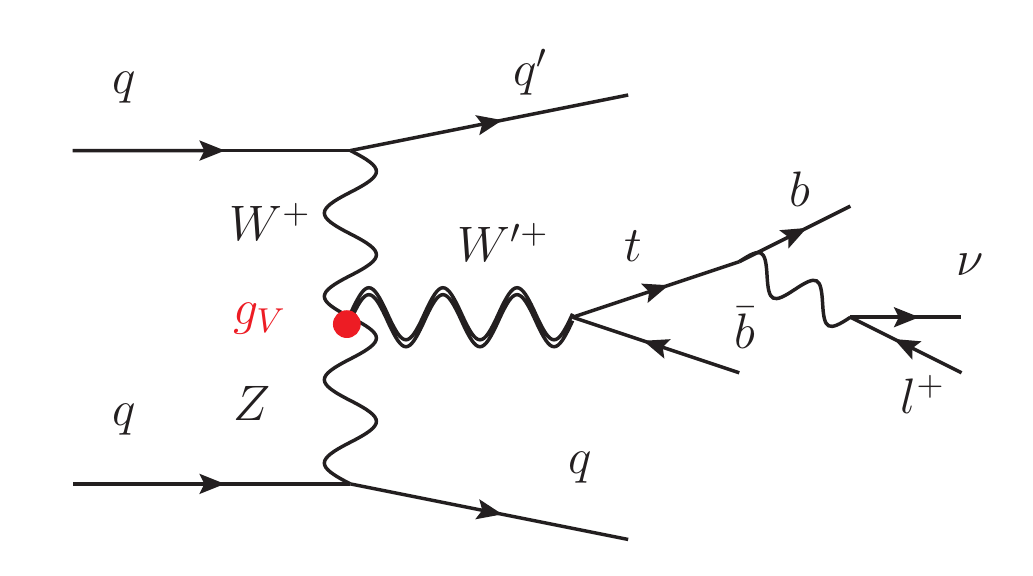} 
\caption{\small The VBF $W^{'} \to tb$ signal. We will consider both the $W^{'+}$ and the $W^{'-}$ processes. }
\label{fig:feyn}
\end{figure}

In this section we 
present an analysis of the sensitivities of future pp colliders, the 14 TeV LHC (LHC-14) and a futuristic 100 TeV collider, on a $W^{'}$ resonance produced by VBF and decaying to $tb$ in the leptonic channel; the corresponding Feynman diagram is shown in Fig. \ref{fig:feyn}. We will consider a $W^{'}$ mass range starting from 1 TeV for the 14 TeV LHC and from 2 TeV for the 100 TeV collider.\\

We generate signal and background events at leading order with MADGRAPH 5 \cite{Alwall:2011uj}. We implement the vector resonance model of sec. \ref{sec:model} in MADGRAPH by using FEYNRULES \cite{Christensen:2008py}. We use the cetq6l1 PDF set \cite{Nadolsky:2008zw} \footnote{We use the dynamical factorization and renormalization scale choice of MADGRAPH \cite{Alwall:2011uj}.}. The events are then passed to PYTHIA 6.4 \cite{Sjostrand:2006za} (with the default tune) for showering and hadronization. Jets are reconstructed with FASTJET \cite{Cacciari:2011ma} by an anti-kt algorithm with cone size $R=$0.4. In order to mimic detector effects we also apply a Gaussian smearing to the jet energy with:
\begin{equation}
\frac{\sigma(E)}{E} = C + \frac{N}{E} + \frac{S}{\sqrt{E}}
\end{equation}
where $E$ is in GeV and $C=0.025$, $N=1.7$, $S=0.58$ \cite{Kulchitsky:2000gg}. The jet momentum is then rescaled by a factor $E^{smeared}/E$.\\

\begin{figure}
\centering
\includegraphics[width=0.6\textwidth]{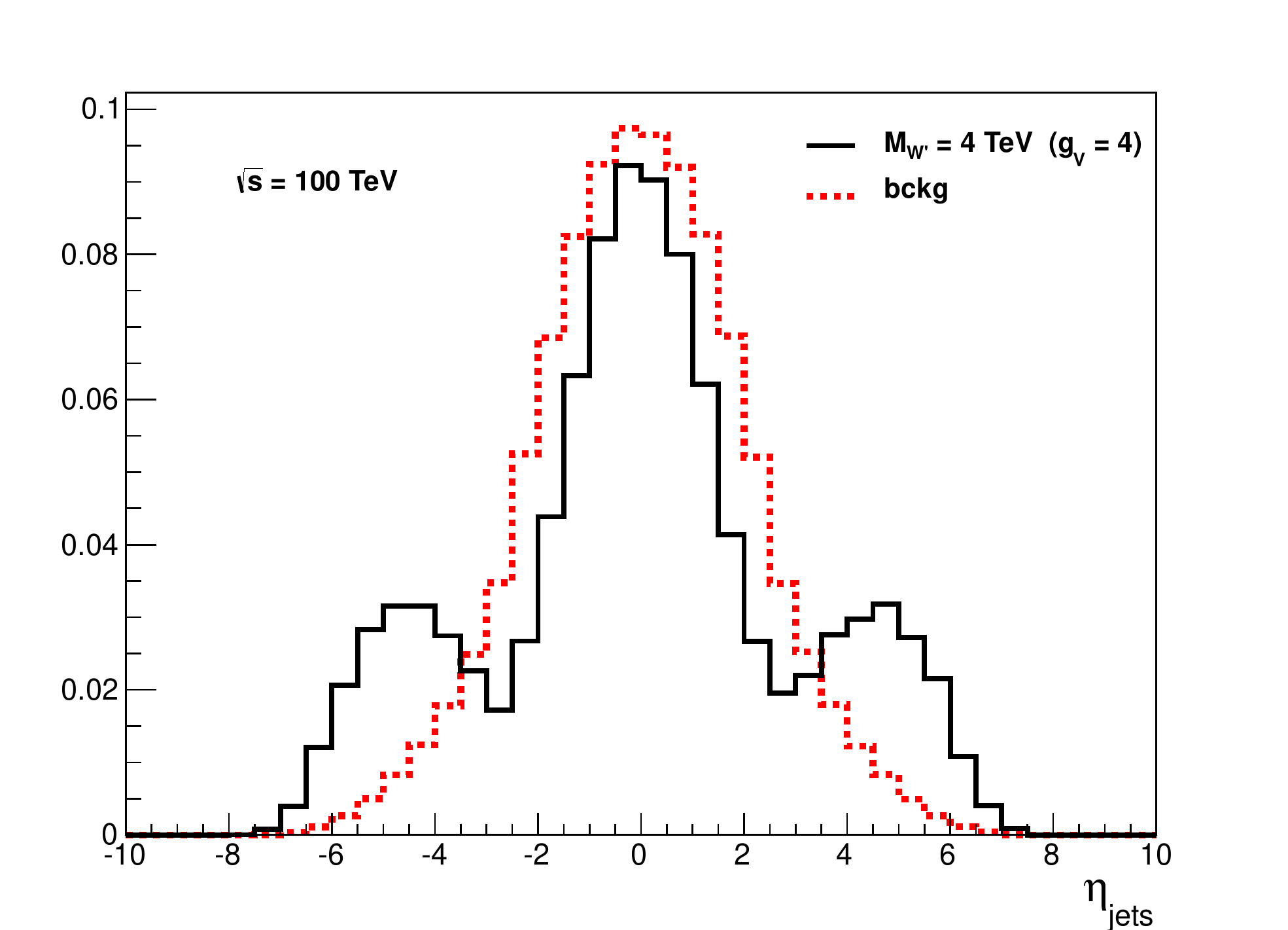} 
\caption{\small (Normalized) rapidity distribution of all of the final jets which have passed the acceptance requirements in eq. (\ref{eq:accept}), with the exception of the $|\eta_j|$ restriction, for the total background (red dashed curve) and the signal with $m_{W^{'}}=4$ TeV, $g_V=4$ (black curve) at a 100 TeV pp collider. }
\label{fig:eta_jets_SB}
\end{figure}

We consider the following final state: exactly one lepton (electron or muon) and at least four jets, of which two must be b-tagged jets:

\begin{equation}
e/ \mu + n_{jet}\ \text{jets} , \qquad  n_{jet} \geq 4 \ (2\ \text{b-tag}).
\label{eq:final-state}
\end{equation}
\noindent
We consider a 0.7 efficiency for the b-tag and a misidentification rate of 0.007 for light jets (0.2 for c-jets). We also require the b-tagged jets to be central, $|\eta_b|<2.5$ \cite{ATLAS:2012ima}.\\ 
We apply slightly different isolation criteria and $p_T$ requirements on the lepton and jets for the analyses at LHC-14 and at a 100 TeV pp collider.

\begin{align}\label{eq:accept}
\begin{split}
&\text{LHC-14:} \\
& p_T\  j > 30 \ \text{GeV} \ , \ p_T\  l > 25 \ \text{GeV} \ , \ \Delta R(l-j)> 0.3 \ , \ |\eta_j|<5 \\[0.2 cm]
& \text{100 TeV:} \\
& p_T\ j > 30 \ \text{GeV}\ , \ p_T\  l > 40 \ \text{GeV}\ , \ \Delta R(l-j)> 0.2 \ , \ |\eta_j|<5, 6  
\end{split}
\end{align}
For the 100 TeV case, we explore a high $W^{'}$ mass region, where the top is boosted and, as a consequence, the lepton tends to be harder and at a lower $R$ separation from the b-jet, which also comes from the top decay. We thus demand a 
harder lepton in order to have a better isolation from the b-jet \cite{ATLAS-CONF-2014-058}. \\
At the 100 TeV collider, the signal is more boosted and, for a significant fraction of the events, the two final forward-backward jets have a rapidity larger than 6, as shown in Fig. \ref{fig:eta_jets_SB}. We thus find advantageous to extend the rapidity acceptance of a future pp collider up to 6. We will present our results for both the rapidity acceptances 
$|\eta_j|<5,6$. \\

The relevant backgrounds to our signal include the $WWbb$, which is mainly made of $t\bar{t}$ events with a minor contribution from single-top $Wt$ events, the $Wbb$+jets and the t-channel single top $tb$+jets \footnote{If we generate the single-top $tb$+jets background with 4 final partons in MADGRAPH, we encounter a non-physical enhancement of the cross section in the region $|\eta_j|\gtrsim$5, which becomes particularly significant at the 100 TeV collider, caused by the emission of a final state gluon collinear with the incoming proton. For example, At the 100 TeV pp collider we obtain a parton level cross section, after acceptance cuts, for $tb$+2 jets which is about 3 times larger than that of $tb$+1 jet. To remove this non-physical effect 
and obtain a reliable estimate of the $tb$+jets background, we produce our t-channel single-top samples by simulating the inclusive $tb$+1 parton process in MADGRAPH (in the four-flavour scheme), which does not contain the extra-radiated collinear gluon. 
In principle, the collinear divergence can be removed by including next-to-leading-order (NLO) virtual corrections. To the best of our knowledge, however, there are currently no available Monte Carlo codes to generate the $tb$ background with four final jets at NLO. All of the other backgrounds, which are not affected by the non-physical collinear enhancement, have been generated with 4 final partons in MADGRAPH. The backgrounds are then passed to PYTHIA for showering and hadronization.   
}. This latter, which has a t-channel topology similar to the signal, represents the dominant background after applying our selection. The $W+$jets background becomes negligible after b-tagging.\\

\begin{figure}
\centering
\includegraphics[width=0.45\textwidth]{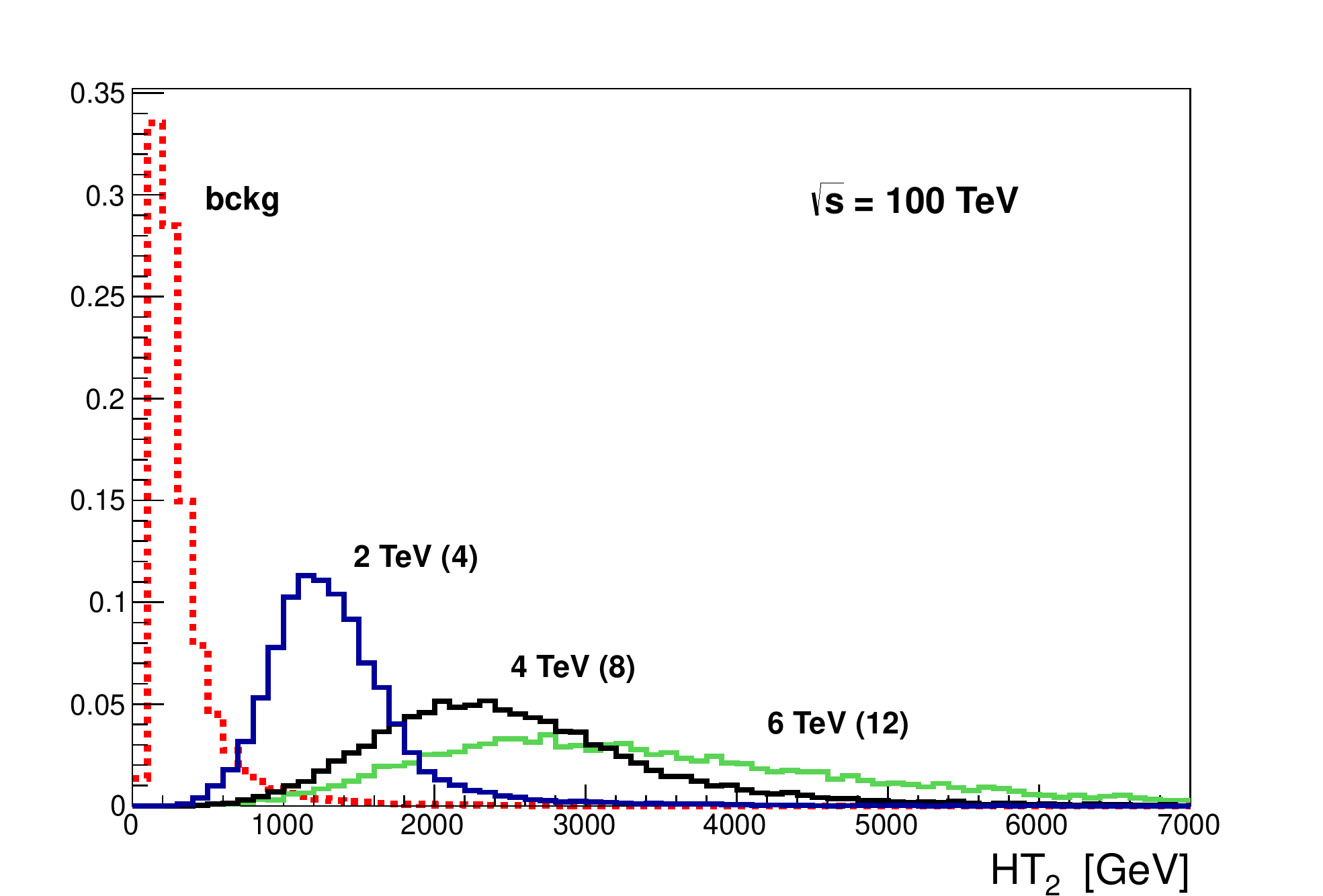} 
\includegraphics[width=0.45\textwidth]{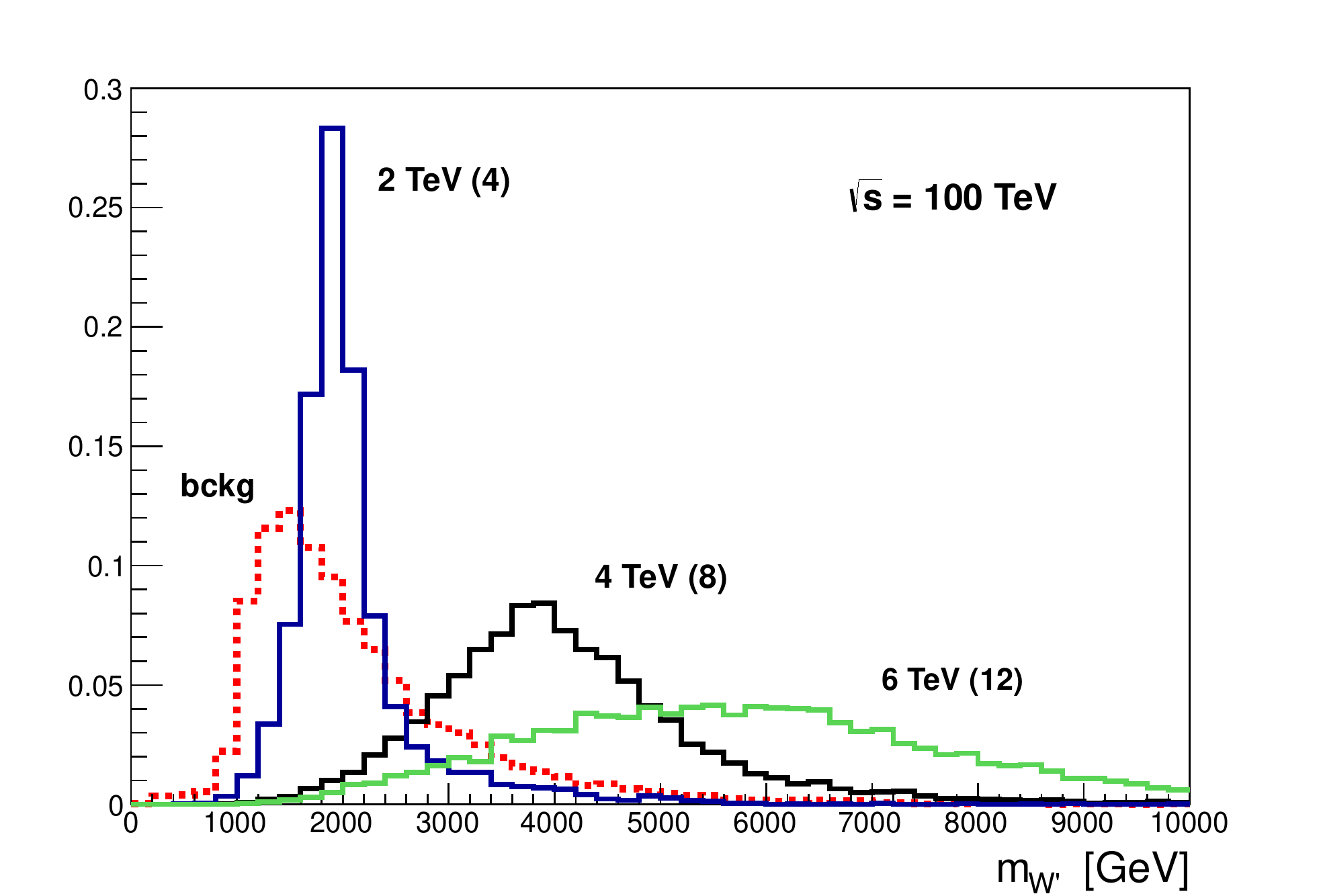} \\
\includegraphics[width=0.45\textwidth]{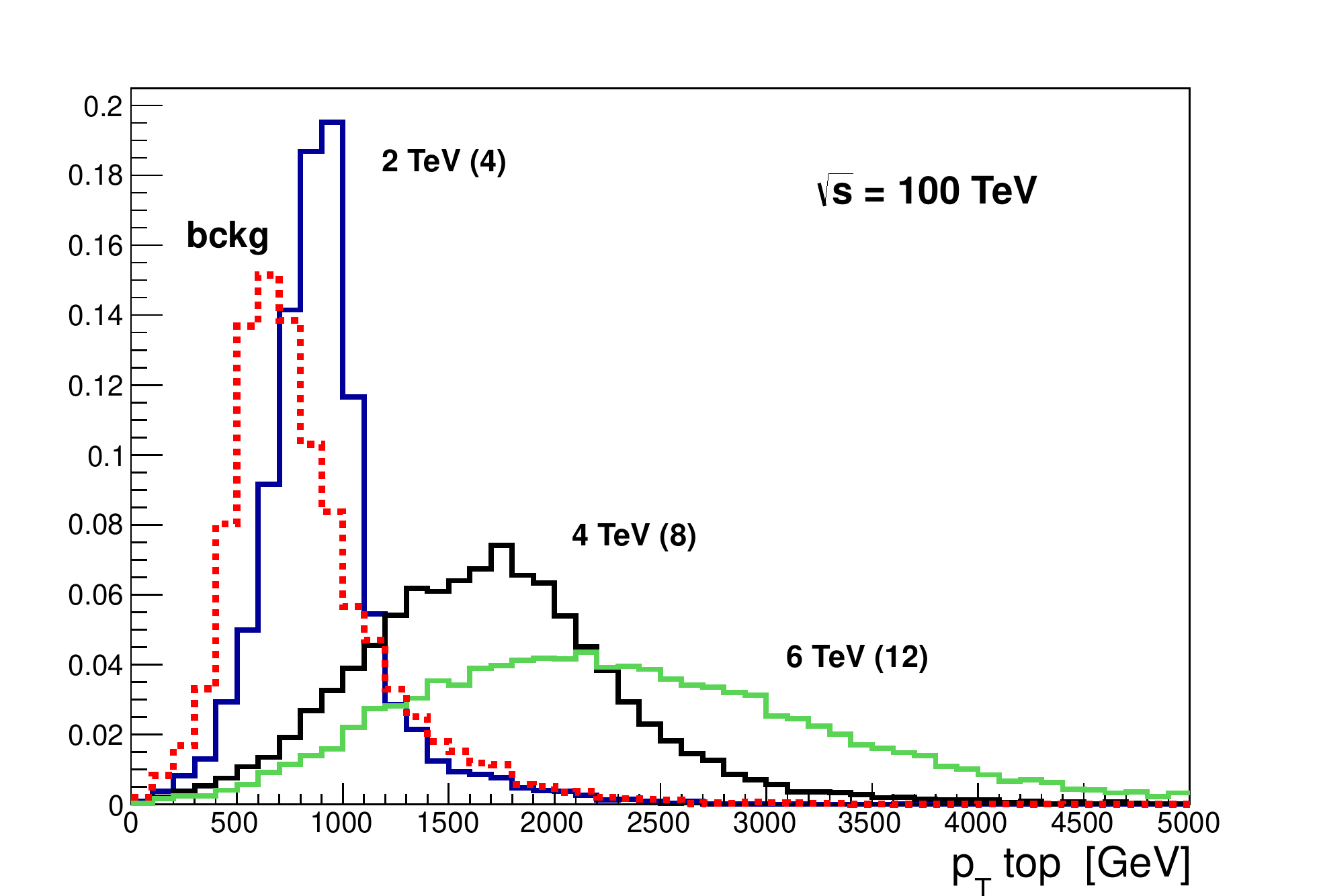} 
\includegraphics[width=0.45\textwidth]{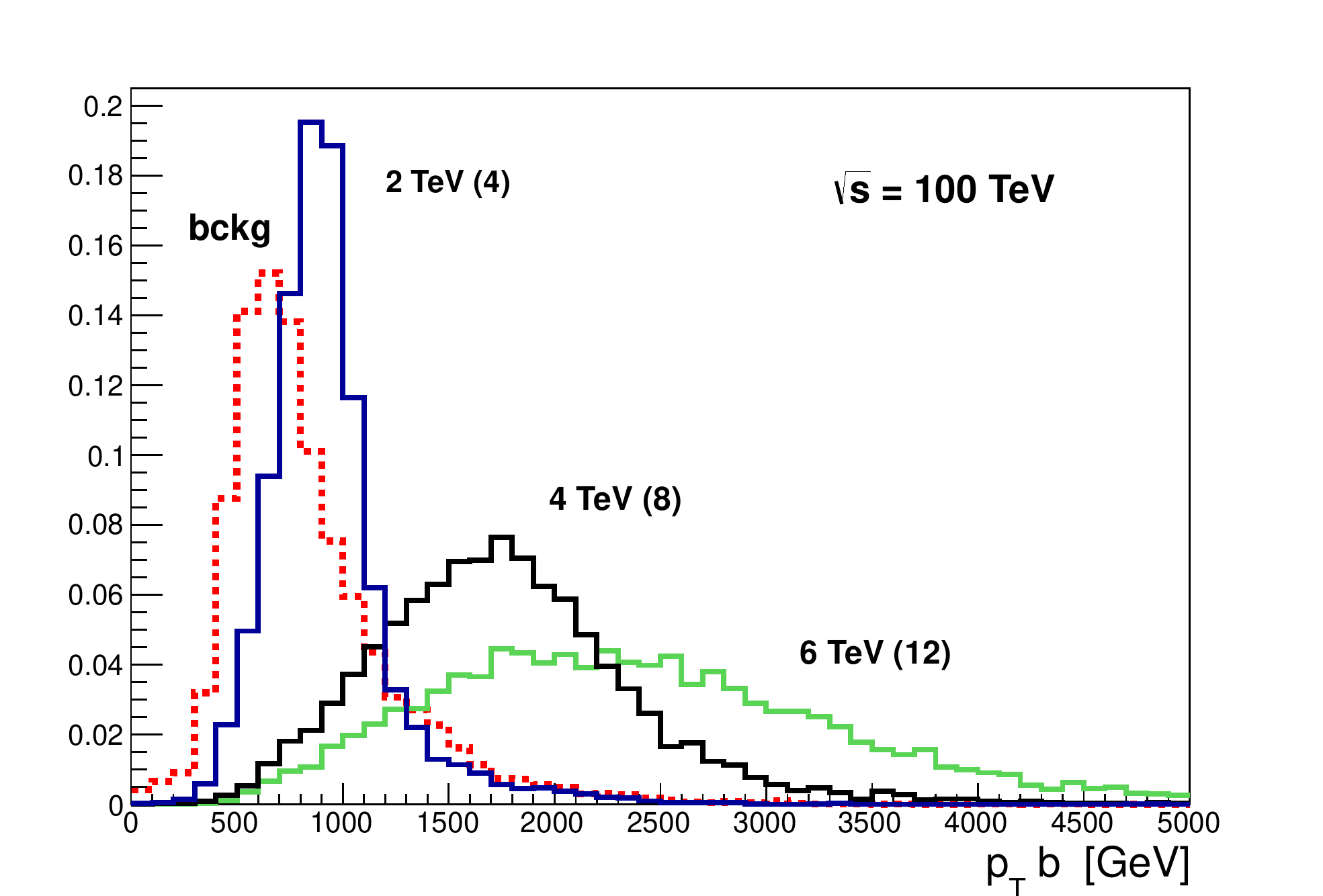} 
\caption{\small Normalized distributions for the total background and the signal for different $W^{'}$ masses and couplings, ($M_{W^{'}}$, $g_V$)= (2 TeV, 4), (4 TeV, 8), (6 TeV, 12) at a future 100 TeV collider. Upper left plot: $HT_2$ distribution, obtained after the acceptance cuts. Upper right plot: $W^{'}$ invariant mass distribution. Lower right (left) plot: $p_T$ distribution of the top (bottom) coming from the $W^{'}$ decays. The $m_{W^{'}}$, $p_T$ top and $p_T$ b distributions have been obtained after the complete selection, except the cuts on (\ref{tab:cut-final}). All of the distributions are shown for the jet rapidity acceptance $|\eta_j|<6$. }
\label{fig:distrib}
\end{figure}

We apply a simple search strategy which relies on the main characteristics of the signal: the presence of a heavy resonance which leads to hard final states and the peculiar VBF topology with the two final forward-backward jets emitted at high rapidity and with a large $\eta$ separation. \\
As a first step of the analysis, we thus impose a cut on $HT_2$, defined as the scalar sum on the $p_T$ of the leading and second-leading jet, which retains at least the 95\% of the signal events for the $W^{'}$ masses analyzed in this study and which already reduces significantly the background. The first plot in Fig. \ref{fig:distrib} shows the $HT_2$ distribution for the background and the signal for several mass-coupling values at the 100 TeV pp collider. We choose the following cut values:

\begin{equation}\label{eq:HT2}
HT_2 > 400 \ \text{GeV} \ [\text{LHC-14}] \qquad HT_2 > 800 \ \text{GeV} \ [\text{100 \ TeV}] 
\end{equation}
We then impose a forward-backward jet tagging and we identify the forward-backward jets (FJ, BJ) in the signal with the following procedure: We require that at least one signal jet must have $\eta>2.5$ and at least one jet $\eta<-2.5$. If more than one jet fulfill the forward-backward requirements, the hardest jet in the forward (backward) $\eta>2.5$ ($\eta<-2.5$) region is identified with the signal forward (backward) jet, FJ (BJ). At this point we also impose a constraint on the rapidity separation between FJ and BJ:

\begin{equation}\label{eq:deltaEta}
|\Delta\eta \ _{ FJ,BJ}|>6 \ \ [\text{LHC-14}] \qquad |\Delta\eta \ _{ FJ,BJ}|>8\ \ [\text{100 \ TeV}] 
\end{equation}
The two hardest jet in the central region $|\eta|<2.5$ are identified with the two final b-jet (we discard the event if less than two signal jets are in the central region). We then reconstruct the neutrino momentum and the top in order to fully reconstruct the $W^{'}$ resonance. The neutrino transverse momentum is reconstructed from a zero total transverse momentum hypothesis. The $p_z$ component is reconstructed from the condition that the neutrino plus the lepton invariant mass, $m(l+\nu)$, gives the $W$ mass and from the top reconstruction procedure. The equation $m^2 (l +\nu)=M^2_W$ gives two $p_z$ solutions\footnote{
In the case of imaginary solutions, we do not reconstruct the neutrino $p_z$ and we fix it to zero during the top reconstruction procedure.}
. One of the two is selected through the top reconstruction procedure. In order to reconstruct the top, the four different $W$ plus b-jet combinations, resulting from the two b-jet identified particles and the two $W$ associated with the two neutrino solutions, are considered. The one which gives the $W$ plus b-jet invariant mass closest to the top mass is selected as the combination of the top decay products. We thus fully reconstruct the neutrino, the top, and we are able to distinguish the b-jet coming from the top from the b-jet directly produced by the $W^{'}$ decay. The $W^{'}$ is finally reconstructed from its decay products, the reconstructed top and the identified bottom. The reconstructed $W^{'}$ invariant mass distribution and the $p_T$ distributions of the top and of the bottom (coming from the $W^{'}$) for the background and for the signal with different $W^{'}$ masses and $g_V$ couplings are shown in Fig. \ref{fig:distrib} for a 100 TeV collider.\\

Once reconstructed the $W^{'}$ and its decay products, the top and the bottom, we impose a bound on the reconstructed $W^{'}$ invariant mass, $m_{W^{'}}$, and on the $p_T$ of the top and of the bottom. The values of the cuts applied for the different $W^{'}$ masses 
are the following for LHC-14: 
 
\begin{equation}
\label{tab:cut-final}
\begin{tabular}[]{c|cccc}
$M_{W^{'}}$ (TeV) & 1 & 1.5 & 2 & 2.5  \\
 \hline
  && &&  \\[-0.3cm]
$m_{W^{'}}\ >$  (TeV) & 0.8 &  1.1 & 1.5 & 1.6 \\
$p_{T} \ b, t \ >$  (TeV) & 0.3 & 0.4 & 0.7 & 0.8 
\end{tabular}
\end{equation} \\
and for a 100 TeV pp collider:
 
\begin{equation}
\label{tab:cut-final}
\begin{tabular}[]{c|ccccc}
$M_{W^{'}}$ (TeV) & 2 & 3 & 4 & 5 & 6 \\
 \hline
  && && & \\[-0.3cm]
$m_{W^{'}}\ >$  (TeV) & 1.5 & 2.5 & 3.5 & 4.0 & 5.0 \\
$p_{T} \ b, t \ >$  (TeV) & 0.75 & 0.9 & 1.5 & 1.5 & 1.5
\end{tabular}
\end{equation}
Tab. \ref{tab:cut-flow-100} and \ref{tab:cut2-flow-100} for the 100 TeV collider and Tab. \ref{tab:cut-flow-14} for the 14 TeV LHC, list the values of the cross section for the signal with different $W^{'}$ masses and $g_V$ couplings and for the backgrounds at each step of the selection.   

\begin{table}[]
\begin{center}
\begin{tabular}[]{c|cccc}
\multicolumn{1}{c|}{{\sf 100 TeV}} & \multicolumn{2}{c}{Acceptance} &  \multicolumn{2}{c}{$HT_{2}>$ 800 GeV} \\[0.1cm]
  & $|\eta_j|<5$ & $|\eta_j|<6$ & $|\eta_j|<5$ & $|\eta_j|<6$ \\[0.1cm]
 \hline
 && && \\[-0.3cm]
 ($M_{W^{'}}$ (TeV), $g_V$) & & & &\\[0.1cm]
&&&&  \\[-0.2cm]
(2, 4) & 2.6 & 3.5 & 2.4 & 3.3 \\ [0.1cm]
(3, 4) & 0.42 & 0.60 & 0.42 & 0.60 \\ [0.1cm]
(4, 4) & 0.095 & 0.14 & 0.095 & 0.14 \\[0.1cm]
(4, 8) & 0.30 & 0.43 & 0.30 & 0.43 \\[0.1cm]
(5, 8) & 0.080 & 0.11 & 0.080 & 0.11 \\[0.1cm]
(6, 12) & 0.036 & 0.051 & 0.036 & 0.051 \\[0.1cm]
\hline
 &&&&  \\[-0.3cm]
WWbb & 3.6 $\cdot 10^5$ & 3.7 $\cdot 10^5$ & 1400 & 1400 \\[0.1cm]
tb+\text{jets} & 1.5 $\cdot 10^4$ & 1.7 $\cdot 10^4$ & 1500 & 1700 \\[0.1cm]
Wbb+\text{jets} & 1.6 $\cdot 10^4$ & 1.6 $\cdot 10^4$ & 1000 & 1000  \\[0.1cm]
&& && \\[-0.2cm]
Tot. BCKG & 3.9 $\cdot 10^5$ & 4.0 $\cdot 10^5$ & 3900 & 4100 \\
\hline  
\end{tabular}
\caption{ \small Signal and background cross sections, in fb, at a 100 TeV pp collider, after the acceptance cuts (\ref{eq:accept}) and the $HT_2$ requirement (\ref{eq:HT2}). Results are shown for two different rapidity acceptances $|\eta_j|<5,6$.
\label{tab:cut-flow-100}}
\end{center}
\end{table}

\begin{table}[]
\begin{center}
\begin{tabular}[]{c|cccc}
\multicolumn{1}{c|}{{\sf 100 TeV}} & \multicolumn{2}{c}{FJ, BJ tag} &  \multicolumn{2}{c}{$|\Delta\eta \ _{ FJ,BJ}|>$8} \\[0.1cm]
  & $|\eta_j|<5$ & $|\eta_j|<6$ & $|\eta_j|<5$ & $|\eta_j|<6$ \\[0.1cm]
 \hline
 && && \\[-0.3cm]
 ($M_{W^{'}}$ (TeV), $g_V$) & & & &\\[0.1cm]
&&&&  \\[-0.2cm]
(2, 4) & 1.7 & 2.5 & 0.95 & 1.7\\ [0.1cm]
(3, 4) & 0.33 & 0.48 & 0.21 & 0.35 \\ [0.1cm]
(4, 4) & 0.077 & 0.11 & 0.053 & 0.088 \\[0.1cm]
(4, 8) & 0.24 & 0.36 & 0.17 & 0.29 \\[0.1cm]
(5, 8) & 0.064 & 0.094 & 0.047 & 0.080 \\[0.1cm]
(6, 12) & 0.030 & 0.044 & 0.023 & 0.036 \\[0.1cm]
\hline
 &&&&  \\[-0.3cm]
WWbb & 89 & 92 & 19 & 21 \\[0.1cm]
tb+\text{jets} & 490 & 590 & 190 & 270 \\[0.1cm]
Wbb+\text{jets} & 140 & 150 & 19 & 22  \\[0.1cm]
&& && \\[-0.2cm]
Tot. BCKG & 720  & 830 & 230 & 310 \\
\hline  
\end{tabular}
\caption{ \small Signal and background cross sections, in fb, at a 100 TeV pp collider after the forward-backward jet tagging and the $|\Delta\eta \ _{ FJ,BJ}|$ restriction (\ref{eq:deltaEta}).  Results are shown for two different rapidity acceptances $|\eta_j|<5,6$. 
\label{tab:cut2-flow-100}}
\end{center}
\end{table}

\begin{table}[]
\begin{center}
\begin{tabular}[]{c|cccc}
{\sf LHC-14} & Acceptance & $HT_{2}>$ 400 GeV & FJ, BJ tag & $|\Delta\eta \ _{ FJ,BJ}|>$6 \\[0.1cm]
 \hline
 && & &  \\[-0.3cm]
 ($M_{W^{'}}$ (TeV), $g_V$) & & &&\\[0.1cm]
&& && \\[-0.2cm]
(1.0, 4) & 0.74 & 0.70 & 0.27 & 0.22 \\ [0.1cm]
(1.5, 4) & 0.088 & 0.088 & 0.046 & 0.039  \\[0.1cm]
(2.0, 8) & 0.043 & 0.043 & 0.025 & 0.022\\[0.1cm]
(2.5, 8) & 0.0091 & 0.0091 & 0.0051 & 0.0044 \\[0.1cm]
\hline
 &&&&  \\[-0.3cm]
WWbb & 13000 & 520 & 7.4 & 4.7 \\[0.1cm]
tb+\text{jets} & 660 & 80 & 9.0 & 5.9 \\[0.1cm]
Wbb+\text{jets} & 680 & 95 & 2.5 & 1.3 \\[0.1cm]
&&&&  \\[-0.2cm]
Tot. BCKG & 14000 & 700 & 19 & 12 \\
\hline  
\end{tabular}
\caption{ \small Signal and background cross sections, in fb, at LHC-14, after the acceptance cuts (\ref{eq:accept}), the $HT_2$ requirement (\ref{eq:HT2}), the forward-backward jet tagging and the $|\Delta\eta \ _{ FJ,BJ}|$ restriction (\ref{eq:deltaEta}).
\label{tab:cut-flow-14}}
\end{center}
\end{table}

\section{Discovery and exclusion reach for the 14 TeV LHC and for a 100 TeV pp collider}\label{sec:results}

\begin{table}[]
\begin{center}
\begin{tabular}[]{c|cc}
\multicolumn{1}{c|}{{\sf LHC-14 }} & \multicolumn{1}{c}{signal} &  \multicolumn{1}{c}{bckg} \\[0.1cm]
\hline
 &&  \\[-0.3cm]
 ($M_{W^{'}}$ (TeV), $g_V$) & & \\[0.1cm]
&& \\[-0.2cm]
(1.0, 4) & 0.17 & 3.1 \\ [0.1cm]
(1.5, 4) & 0.030 & 0.90  \\ [0.1cm]
(2.0, 8) & 0.012 & 0.10  \\[0.1cm]
(2.5, 8) & 0.0025 & 0.063 \\[0.1cm]
\hline  
\end{tabular}
\caption{ \small Signal and background cross sections, in fb, at LHC-14 after the complete selection.
\label{tab:final-14}}
\end{center}
\end{table}

\begin{table}[]
\begin{center}
\begin{tabular}[]{c|cccc}
\multicolumn{1}{c|}{{\sf 100 TeV}} & \multicolumn{2}{c}{signal} &  \multicolumn{2}{c}{bckg} \\[0.1cm]
  & $|\eta_j|<5$ & $|\eta_j|<6$ & $|\eta_j|<5$ & $|\eta_j|<6$ \\[0.1cm]
 \hline
 && && \\[-0.3cm]
 ($M_{W^{'}}$ (TeV), $g_V$) & & & &\\[0.1cm]
&&&&  \\[-0.2cm]
(2, 4) & 0.56 & 1.1 & 70 & 100\\ [0.1cm]
(3, 4) & 0.13 & 0.25 & 31 & 45 \\ [0.1cm]
(4, 4) & 0.022 & 0.042 & 4.8 & 7.2 \\[0.1cm]
(4, 8) & 0.082 & 0.15 & 4.8 & 7.2 \\[0.1cm]
(5, 8) & 0.028 & 0.051 & 3.6 & 4.9 \\[0.1cm]
(6, 12) & 0.013 & 0.022 & 1.4 & 1.8 \\[0.1cm]
\hline  
\end{tabular}
\caption{ \small Signal and background cross sections, in fb, at a 100 TeV pp collider after the complete selection.
\label{tab:final-100}}
\end{center}
\end{table}

The final results of our selection are shown on Tab. \ref{tab:final-14} for the 14 TeV LHC and on Tab. \ref{tab:final-100} for a futuristic 100 TeV collider. \\
From the final results in Table \ref{tab:final-14} and \ref{tab:final-100}  we are able to estimate the discovery/exclusion reach in the $W^{'}$ (mass, coupling) parameter space \footnote{
We set a 95$\%$ C.L. exclusion limit if the goodness-of-fit test of the signal plus background hypothesis with Poisson distribution gives a p-value less than 0.05 and we claim a 5$\sigma$ discovery if the p-value of the SM-only hypothesis is less than 2.8$\cdot$10$^{-7}$.}.
To do this, we consider a scaling of the signal cross section as $g^2_V$ with the coupling, and, for the parameter space at small couplings, $g_V \lesssim 4$, we also include 
the variation of the signal cross section with $BR(W^{'}\to tb)$, which, for $g_V \lesssim 4$, begins to change significantly with the coupling, as shown in Fig. \ref{fig:BR-width}. \\

\begin{figure}
\centering
\includegraphics[width=0.6\textwidth]{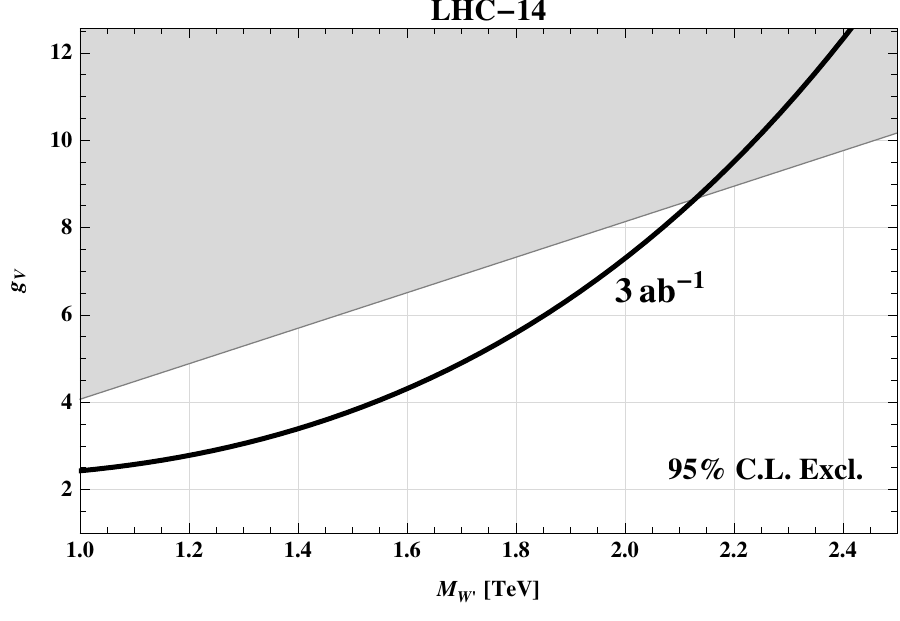} 
\caption{\small The 14 TeV LHC exclusion potential, with 3 ab$^{-1}$, on a $W^{'}$ produced via VBF in the $tb$ channel. The shaded area corresponds to values $g_V v/M_{V} > 1$ which are indicative of a theoretically excluded region (where $v/f\gtrsim 1$) in MCHM. }
\label{fig:reach-14}
\end{figure}

\begin{figure}
\centering
\includegraphics[width=0.6\textwidth]{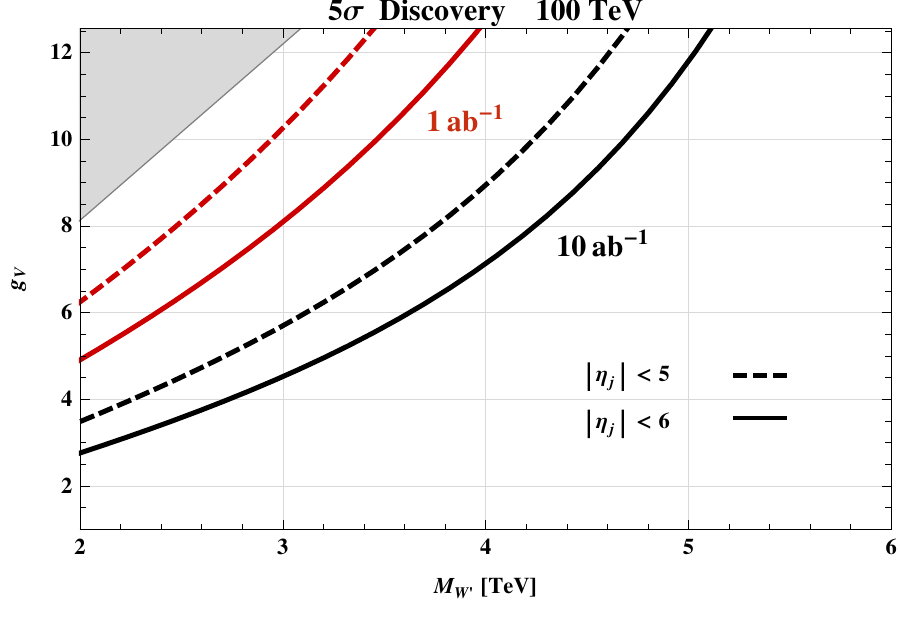} \\[0.4cm]
\includegraphics[width=0.6\textwidth]{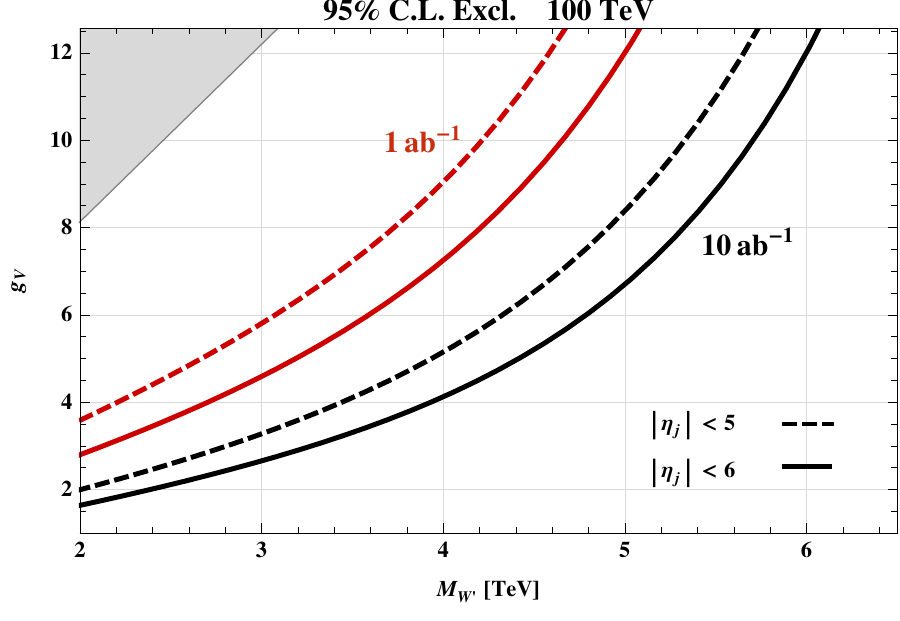} 
\caption{\small The 100 TeV pp collider reach, with 1 and 10 ab$^{-1}$, on a $W^{'}$ produced via VBF in the $tb$ channel. Upper plot: 5$\sigma$ discovery potential. Lower Plot: 95\% CL exclusion reach. The continuous (dotted) curves are obtained for a jet rapidity acceptance $|\eta_j|<$ 6 (5). The shaded areas in the upper-left corner of the parameter space correspond to values $g_V v/M_{V} > 1$ which are indicative of a theoretically excluded region (where $v/f\gtrsim 1$) in MCHM. }
\label{fig:reach-100}
\end{figure}

The final reach of a futuristic 100 TeV pp collider on a $W^{'}$ produced via VBF is shown in Fig. \ref{fig:reach-100}, the  exclusion potential of the high luminosity LHC-14 is presented in Fig. \ref{fig:reach-14}. The discovery/exclusion reach is presented in the $W^{'}$ (mass, coupling) parameter space.
As anticipated in sec. \ref{sec:model}, the shaded region in the upper-left corner of the $W^{'}$ parameter space is indicative of a theoretically forbidden region in MCHM with a pGB Higgs \footnote{We stress again that the constraint $g_V v/M_V<1$ is not strict and is subject to $\mathcal{O}(1)$ corrections.  }.
We find that the 14 TeV LHC can access only a small portion of the MCHM parameter space. The high-luminosity LHC, with 3 ab$^{-1}$ can exclude\footnote{The discovery reach of the 14 TeV LHC does not cover the theoretically allowed region.} a $W^{'}$ vector resonance up to about 2.1 TeV. This mass range for the $W^{'}$ is quite in tension with the electroweak-precision-tests, since the S parameter gives a lower bound of about 2 TeV on the $W^{'}$ mass \cite{Pappadopulo:2014qza, Ciuchini:2013pca}. But it would be nevertheless important to have a complementary information through a direct measurement at LHC-14.
A futuristic 100 TeV pp collider has a much wider sensitivity. The upper plot in Fig. \ref{fig:reach-100} shows that a futuristic 100 TeV collider can discover, at 5$\sigma$, a $W^{'}$ in the VBF channel with masses up to 5.1 (4) TeV with 10 (1) ab$^{-1}$ of integrated luminosity in the large $g_V$ coupling region. The exclusion potential of the 100 TeV collider, as shown in the lower plot of Fig. \ref{fig:reach-100}, extends up to $W^{'}$ masses of 6.1 (5.1) TeV with 10 (1) ab$^{-1}$. These values refer to a jet-rapidity acceptance $|\eta_j|<6$. We find that the reach of a future 100 TeV pp collider is significantly enhanced, by about a 10\% in the $W^{'}$ mass reach, if the rapidity acceptance on the jets can be increased from 5, the present LHC-14 acceptance, up to 6. 

\section{Conclusions}\label{sec:conclusions}

In this study we have presented a first estimate of the reach of future pp colliders, the 14 TeV LHC and a futuristic 100 TeV pp collider, on a vector resonance, specifically a $W^{'}$, produced via VBF, and decaying dominantly into $tb$. The analysis is motivated by Composite Higgs, Randall-Sundrum and Little Higgs scenarios, which predict the existence of vector resonances with a large coupling to $W$ and $Z$ longitudinal bosons.
In particular, in MCHM with partial compositeness, the standard DY production channel is suppressed at large coupling while the VBF production is enhanced and could thus provide a unique opportunity to directly test the large-coupling regime of the theory.\\
We have derived a search strategy for the $W^{'}$ produced by VBF and decaying to $tb$ and obtained the estimated reach of the 14 TeV LHC and of a 100 TeV pp collider on the (mass, coupling) $W^{'}$ parameter space. Our results are shown in  Fig. \ref{fig:reach-14} for LHC-14 and in Fig. \ref{fig:reach-100} for the 100 TeV collider. We find that, due to a low VBF production cross section at $\sqrt{s}=14$ TeV, the LHC-14 can access only a small portion of the $W^{'}$ parameter space, with the possibility to exclude a $W^{'}$ vector resonance up to about 2.1 TeV with 3 ab$^{-1}$ in the large-coupling regime. Although this region of parameter space has some tension with electroweak precision data, an analysis of the VBF channel at LHC-14 could provide a direct complementary confirmation of the exclusion. Fig. \ref{fig:reach-100} shows that the sensitivity of a future 100 TeV collider on a $W^{'}$ produced via VBF is high. The discovery reach, at 5$\sigma$, extends up to $W^{'}$ masses of 5.1 (4) TeV with 10 (1) ab$^{-1}$ of integrated luminosity in the large-coupling region. While a future 100 TeV collider can exclude a $W^{'}$ in the large-coupling regime, with masses up to 6.1 (5.1) TeV with 10 (1) ab$^{-1}$. We finally find that the 100 TeV collider reach is considerably increased for a jet rapidity acceptance extended up to $|\eta_j|<6$. 

\section*{Acknowledgments}
This material is based upon work supported by the National Science Foundation under Grant No. PHY-0854889.

\end{document}